\documentclass[twocolumn,english,aps,superscriptaddress,showpacs]{revtex4-1}
\usepackage[T1]{fontenc}
\usepackage{xcolor,graphicx}
\usepackage[colorlinks=true]{hyperref}

\usepackage{subfigure}
\usepackage{amssymb}
\usepackage{verbatim}
\usepackage{amsmath}
\usepackage{amscd}
\usepackage{amssymb}
\usepackage{setspace}
\usepackage{amsfonts}

\usepackage{amsmath,amsthm,amssymb}
\usepackage{setspace}
\usepackage{pgf,tikz}
\usetikzlibrary{shapes,snakes,calendar,matrix,backgrounds,folding}

\newcommand{\ket}[1]{\ensuremath{|#1\rangle}}
\newcommand{\bra}[1]{\ensuremath{\langle#1|}}

\newcommand{\tr}{\mathop{\rm tr}\nolimits}
\newcommand{\Tr}{\mathop{\rm Tr}\nolimits}
\newcommand{\argmax}{\mathop{\rm argmax}\nolimits}

\newcommand{\trho}{\tilde{\rho}}

\newcommand\ip[2]{\ensuremath{\langle#1,#2\rangle}}
\newcommand\range{\mathop{\rm range}\nolimits}

\newcommand{\nc}{\newcommand}
\nc{\ketbra}[2]{|#1\rangle\!\langle#2|}
\nc{\braket}[2]{\langle#1|#2\rangle}

\newtheorem{lemma}{Lemma}

\newtheorem{obs}{Observation}

\def\D{\mathcal{D}}

\def\K{\mathcal{K}}
\def\P{\mathcal{P}}

\makeatother

\begin{document}
%% End-Of-Header

\title{From Ground States to Local Hamiltonians}

\author{Jianxin Chen}
\affiliation{Department of Mathematics \&
  Statistics, University of Guelph, Guelph, Ontario, Canada}
\affiliation{Institute for Quantum Computing, University of Waterloo,
  Waterloo, Ontario, Canada}

\author{Zhengfeng Ji} \affiliation{Institute for Quantum Computing,
  University of Waterloo, Waterloo, Ontario, Canada}
\affiliation{State Key Laboratory of Computer Science, Institute of
  Software, Chinese Academy of Sciences, Beijing, China}

\author{Bei Zeng} 
\affiliation{Department of Mathematics \&
  Statistics, University of Guelph, Guelph, Ontario, Canada}
\affiliation{Institute for Quantum Computing, University of Waterloo,
  Waterloo, Ontario, Canada}

\author{D.~L. Zhou}
\affiliation{Beijing National Laboratory for
  Condensed Matter Physics, and Institute of Physics, Chinese Academy
  of Sciences, Beijing 100190, China}

\begin{abstract}
  Traditional quantum physics solves ground states for a given
  Hamiltonian, while quantum information science asks for the
  existence and construction of certain Hamiltonians for given ground
  states. In practical situations, one would be mainly interested in
  local Hamiltonians with certain interaction patterns, such as
  nearest neighbour interactions on some type of lattices.  
  A necessary condition for a space $V$ to be the ground-state
  space of some local Hamiltonian with a given interaction pattern, is
  that the maximally mixed state supported on $V$ is uniquely
  determined by its reduced density matrices associated with the given
  pattern, based on the principle of maximum entropy. 
  However, it is unclear whether this condition is in general also sufficient. 
  We examine the situations for the existence of such a local 
  Hamiltonian to have $V$ satisfying the necessary condition mentioned above
  as its ground-state space, by linking to faces of the convex
  body of the local reduced states.  We further discuss some
  methods for constructing the corresponding local Hamiltonians with
  given interaction patterns, mainly from physical points of view,
  including constructions related to perturbation methods, local
  frustration-free Hamiltonians, as well as thermodynamical ensembles.
\end{abstract}

\date{\today}


\maketitle

\section{Introduction}

Traditional quantum many-body physics focuses on finding ground state
energy and the corresponding ground states for some given
Hamiltonians.  A naturally-occurring Hamiltonian involves only one and
two-body interactions in most cases. The new field of quantum
information science, however, focuses more on studying quantum
states ~\cite{NC00}. Quantum states are ``information carriers'' of
quantum information, upon which communication is conveyed, and
computation is implemented. After years of development, it becomes
convincing today that quantum communication and computation offers the
possibility of secure and high rate information transmission and fast
computational solution of certain important problems, which is at the
heart of modern information technology.

One major direction of quantum information science is to study
correlations in many-body quantum systems. Here the term correlation
is used instead of entanglement, due to the fact that a quantum state
contains both classical and quantum correlation, which both contribute
to real physical phenomena. Traditionally, correlation is
characterized by correlation functions, which are directly related to
experimental measurements of physical observables.  Quantum
information science brings new angles to study correlations, from
information scientific points of view.

An interesting viewpoint on correlation in quantum states is based on the
principle of maximum entropy, which is advocated by Jaynes in the
study on the foundation of statistical mechanics~\cite{Jay57}.  The
principle says that if an $n$-particle quantum state $\rho$ has the
maximum entropy among all the $n$-particle states with the same
$k$-particle reduced density matrices ($k$-RDMs) as those of $\rho$,
then $\rho$ contains no more information than that contained in its
$k$-RDMs.  And such a $\rho$ consisting with the given $k$-RDMs is
indeed unique.  In this sense $\rho$ contains no irreducible $r$-particle
correlation for any $r>k$~\cite{LW02,Zho08}.

In the case that $\rho$ is a pure state, $\rho$ is uniquely determined by
its $k$-RDMs, based on the principle of maximum entropy. It simply means
that there does not exist any other state, pure or mixed, which has
the same $k$-RDMs as those of $\rho$. Well known examples include,
almost all three-qubit pure states are uniquely determined by their
$2$-RDMs~\cite{LPW02}; almost every pure state of many-body quantum
systems (with equal dimensional subsystems) is uniquely determined by
its RDMs of just over half of the parties~\cite{LW02,JL05}; $W$-type
states are uniquely determined by their $2$-RDMs~\cite{PR08}; and the
only $n$-particle pure states which cannot be determined by their
$(n-1)$-RDMs are those $GHZ$-type states~\cite{WL08}.

A many-body Hamiltonian $H$ is $k$-local if $H=\sum_i H_i$, where each
term $H_i$ acts non-trivially on at most $k$-particles.  In practical
situations, one would be mainly interested in $k$-local Hamiltonians
with certain interaction patterns, such as nearest neighbour
interactions on some type of lattices.  That is, for a given space
$V$, one would like to know whether $V$ can be the ground-state space
of some $k$-local Hamiltonian $H=\sum_i H_i$ which contains only
certain terms of $k$-particle interactions; and if such a $k$-local
Hamiltonian exists, how to find it.

In this paper, we address this question by starting from a natural necessary
condition for a space $V$ to be the ground-state space of some local
Hamiltonian with a given interaction pattern. That is, the maximally
mixed state supported on $V$ is uniquely determined by its reduced
density matrices associated with the given interaction pattern, based
on the principle of maximum entropy. 
This condition builds an interesting link
between correlations of quantum states and ground-state spaces of
local Hamiltonians, unfortunately it is unclear whether this condition is in general
also sufficient.  We examine the situations for the existence of such a local 
Hamiltonian to have $V$ satisfying the necessary condition mentioned above 
as its ground-state space, by linking to faces of the convex
body of the local reduced states. We then further discuss some methods for
constructing a corresponding $k$-local Hamiltonian, mainly from
physical points of view, including constructions related to
perturbation methods, local frustration-free Hamiltonians, as well as
thermodynamical ensembles.

We organize our paper as follows. In Sec.~\ref{sec:rel}, we give a 
formal definition of local Hamiltonians of a given interaction pattern, 
and review the convex geometry viewpoint of their ground-state spaces.
In Sec.~\ref{sec:kcorr}, we introduce the concept of $\K$-correlated subspaces
as link it to the correlation of ground-state spaces of local Hamiltonians and
discuss its meaning in terms of convex geometry. 
In Sec.~\ref{sec:nonexpose}, we examine 
in more detail the situations where $\K$-correlated subspace may fail
to be the ground-state space of the corresponding local Hamiltonian of 
given interaction pattern, and provide a perturbation method to construct
such a Hamiltonian if it exists. In Sec.~\ref{sec:FF}, we provide another method of
finding the local Hamiltonians of some frustrated systems starting
from some frustration-free systems, which, combining with the
perturbation method, succeeds in finding the local Hamiltonians in
certain special cases.  For instance, this allows us to identify
Hamiltonians for almost all three-qubit states, and the $n$-qubit $W$
states with only nearest neighbour interactions on a one-dimensional
spin chain.  In Sec.~\ref{sec:TE}, we provide a general method of
finding the local Hamiltonians from a thermal ensemble idea.  Finally,
a summary and discussion is given in Sec.~\ref{sec:CD}.

\section{Local Hamiltonians and Convex Geometry}
\label{sec:rel}

This section discusses the ground-state space properties of
local Hamiltonians. We start from a formal discussion of local Hamiltonians with 
given interaction patterns.

%\subsection{$\mathcal{K}$-local Hamiltonians}

Consider an $n$-particle system.  We specify a pattern $\cal{K}$,
where each element $K_j \in \cal{K}$ is a subset of $\{1, 2, \ldots,
n\}$ with $|K_j| = k$ (here $|K_j|$ is the size of $K_j$).  A
Hamiltonian $H = \sum_i H_i$ is called $\K$-local if each $H_i$ acts
nontrivially on at most $k$ particles in some $K_j \in \cal{K}$. In
practice, the choice of such a pattern $\K$ is usually related to
certain spacial geometry considerations, such as nearest neighbour
particles with respect to some spin lattices.

As an example, the Hamiltonian $H$ of three qubits
\begin{equation}
  \label{eq:H3}
  H = J (X_1 X_2 + X_2 X_3) + B (Z_1 + Z_2 + Z_3)
\end{equation}
is $\K$-local where
\begin{equation}
  \K=\{\{1, 2\}, \{2, 3\}\}.
\end{equation}
Here $X_j, Y_j, Z_j$ are Pauli $X, Y, Z$ operators acting on the $j$th
qubit.

Note that for any $\tilde{\K} \supseteq \K$, a Hamiltonian $H$ is
$\K$-local is also $\tilde{\K}$-local. Furthermore, for some $k' \ge
k$ and a pattern $\K'$ with $\vert K'_j \vert \le k'$, $H$ is also
$\K'$-local if for any $K_i \in \K$ there exists some $K'_j \in \K'$
such that $K_i \subseteq \K'_j$. For instance, the Hamiltonian given in
Eq.(\ref{eq:H3}) is also $\{\{1, 2\}, \{2, 3\},\{1, 3\}\}$-local or
$\{\{1, 2, 3\}\}$-local.  In practice, we would usually be interested
in the smallest number $k$ and the smallest possible set $\K$ such
that $H$ is $\K$-local.

Let $\D$ be the set of density matrices of $n$-particles. For any
given pattern $\K$, list all the elements $K_i \in \K$ as a vector
$(K_1, K_2, \ldots, K_M)$ in a fixed order, where $M$ is the size of
$\K$.  When $M = {n \choose k}$, ${\K}$ contains all the $k$-element
subset of $\{1, 2, \ldots, n\}$.  Let $\gamma_{K_i}$ be the $k$-RDM of
particles in $K_i \in \K$.  For $\rho \in \D$, let
\begin{equation}
  \vec{R}_{\K}(\rho) = (\gamma_{K_1},\gamma_{K_2},\ldots,\gamma_{K_M}),
\end{equation}
which is a vector whose elements are $k$-RDMs of $\rho$.

%This is because that for any state $\rho$ supported on $V$, $\tr(\rho
%H)$ equals to the ground energy. For any state $\sigma\in
%A_{\K}(\rho)$, $\tr(\sigma H)$ also equals to the ground energy as
%$\K$-local Hamiltonian only ``sees'' the $k$-RDMs of particles in some
%$K_j\in\cal{K}$. Therefore, $\sigma$ is also supported on $V$.

Note a simple fact that the set
\begin{equation}
  \D_{\K} = \{\vec{R}_{\K}(\rho) \,|\, \rho\in\D\}
\end{equation}
is a closed convex set. 
Indeed it has been known that there is a natural connection between ground-state 
spaces to exposed faces of the convex set $\D_{\K}$ (see, for instance, ~\cite{VC06,FNW92}), that we briefly review here.

We first recall some notations from convex analysis. For a convex set
$C$, its dual cone $\P(C)$ is
\begin{equation}
  \P(C) = \left\{ \vec{y} \,|\, \forall\, \vec{x}\in C,
    \ip{\vec{x}}{\vec{}y} \ge 0 \right\}.
\end{equation}
Let the dual cone of $\D_{\K}$ be $\P_{\K}$. For the vectors
$\vec{x}=(\gamma_1,\gamma_2,\ldots,\gamma_M)$ and
$\vec{y}=(H_1,H_2,\ldots,H_M)$ with Hermitian $H_j$s,
$\ip{\vec{x}}{\vec{y}}$ is defined as $\sum_{j=1}^M
\Tr(H_j\gamma_j)$. Any point $\vec{H}$ of the form
$(H_1,H_2,\ldots,H_M)$ defines a $\K$-local Hamiltonian $H =
\sum_{j=1}^M H_j$. Moreover, we have $\ip{\vec{R}_{\K}(\rho)}{\vec{H}}
=\Tr(\rho H)$. This allows us to visualize $\K$-local Hamiltonians as
hyperplanes in the space containing $\D_{\K}$. More specifically, let
$\vec{H}$ be a point that corresponds to the Hamiltonian $H$ and
define a hyperplane also denoted as $H$ to be
\begin{equation*}
  H = \{\vec{x}\,|\, \ip{\vec{x}}{\vec{H}} = 0 \}.
\end{equation*}

For any convex set $C$, a subset $F$ is called a face on $C$ if
\begin{enumerate}
\item $F$ is a convex set, and
\item For any line segment $L\subseteq C$, if $L$ intersects $F$ at
  some point other than the two end points of $L$, then $L\subseteq
  F$~\cite{Roc96}.
\end{enumerate}
A face $F$ is exposed if there exists some element $\vec{y}$ in the dual cone
$\P(C)$ such that $ \ip{\vec{x}}{\vec{}y} = 0,\ \forall\, \vec{x}\in F$ and
$ \ip{\vec{x}}{\vec{}y} > 0,\ \forall\, \vec{x}\notin F$.

Let the set $F_{V}$ be the image in $\D_\K$ for the states supported on the space
$V$. That is,
\begin{equation}
  \label{eq:FV}
  F_{V}=\big\{\vec{R}_\K (\sigma) \, \vert \, \range(\sigma) \subseteq V \big\},
\end{equation}
then for any $V$ that is a ground-state space of some $\K$-local Hamiltonian,
$F_{V}$ is an exposed face of $\D_\K$.

\section{$\mathcal{K}$-correlated spaces}
\label{sec:kcorr}

For any $n$-particle quantum state $\rho$, define a set $A_{\K}(\rho)$
of $n$-particle quantum states which have the same array of $k$-RDMs
as $\rho$, i.e.
\begin{equation}
  A_{\K}(\rho) = \big\{ \sigma \in \D \, \vert \, \vec{R}_{\K}(\sigma) =
  \vec{R}_{\K}(\rho) \big\}.
\end{equation}

%Now for a given $\K$-local Hamiltonian $H$, we observe the following
%property for the ground-state space $V$ of $H$.
%\begin{obs}
%\label{obs:ground}
%  Let $V$ be the ground-state space of a $\K$-local Hamiltonian $H$. 
%  Then for any state $\rho$ supported on $V$, any state $\sigma$ in
%  $A_{\K}(\rho)$ is also supported on $V$.
%\end{obs}

Let $\trho_{\K}$ denote the state of maximum entropy among all the
states in $A_{\K}(\rho)$, i.e.
\begin{equation}
 \trho_{\K} =\argmax\, \big\{ S(\sigma ) \, | \, \sigma \in
  A_{\K}(\rho) \big\},
  \label{eq:tilde}
\end{equation}
where the von Neumann entropy $S(\rho)= -\Tr(\rho
\log\rho)$. Note that $\trho_{\K}$ is indeed unique.

Based on the principle of maximum entropy, $\trho_{\K}$ contains no
more information than that is contained in the reduced density
matrices $\gamma_{K_i}s$.  Therefore, if $\rho = \trho_{\K}$, then
$\rho$ is the state containing no more information, than that is
contained in the reduced density matrices $\gamma_{K_i}s$. In other
words, $\rho$ can be determined without ambiguity from
$\gamma_{K_i}s$.  In this sense, we say that the state $\rho$ is
uniquely determined by $\gamma_{K_i}s$, and call it
$\K$-correlated. That is, an $n$-particle state $\rho$ is called
$\K$-correlated if $\rho=\trho_{\K}$.

As an example for $\K$-correlated states, consider the three-qubit
state
\begin{equation}
  \label{eq:rhoc}
  \rho_c=\frac {1} {2} ( \vert 000\rangle \langle 000\vert + \vert
  111\rangle \langle 111\vert ).
\end{equation}
For $\K=\{\{1,2\},\{2,3\}\}$, it is straightforward to check that
among all the three qubit states with the same $2$-RDMs for particles
$\{1,2\}$ and $\{2,3\}$, $\rho_c$ has the maximum entropy. So $\rho_c$
is $\K$-correlated.  On the other hand, the three-qubit $GHZ$ state
\begin{equation}
\label{eq:GHZ}
\vert GHZ\rangle =\frac{1}{\sqrt{2}}(\vert 000\rangle +\vert 111\rangle)
\end{equation}
has the same $2$-RDMs for particles $\{1,2\}$ and $\{2,3\}$ as
$\rho_c$, but $\rho_c$ has a larger von Neumann entropy than that of
$\ket{GHZ}$. Therefore, $\ket{GHZ}$ is not $\K$-correlated.

Note that similar as the case of $\K$-local Hamiltonians, for any
$\tilde{\K}\supseteq\K$, a state $\rho$ is $\K$-correlated is also
$\tilde{\K}$-correlated. Furthermore, for some $k'>k$ and a pattern
$\K'$ with $|K'_j|=k'$, $\rho$ is also $\K'$-correlated if for any
$K_i\in\K$ there exists some $K'_j\in\K'$ such that
$K_i\subseteq\K'_j$. In practice, for a given $\rho$, we usually would
like to find the smallest possible number $k$ and the smallest
possible set $\K$ such that $\rho$ is $\K$-correlated.

For a space $V$, if the maximally mixed state $\rho_V$ supported on
$V$ is $\K$-correlated, then we call the space $\K$-correlated. 
The following simple observation then links the ground-state space of $\K$-local
Hamiltonians and $\K$-correlated space.
\begin{obs}
  \label{ob:rel}
  If $V$ is the ground-state space
  of some $\K$-local Hamiltonian, then $V$ is $\K$-correlated.
\end{obs}

This is because that for any state $\rho$ supported on $V$, 
$\tr(\rho H)$ equals to the ground energy. Then obviously 
the maximally mixed state $\rho_V$ supported on
$V$ has the maximum entropy among all states in 
$A_{\K}(\rho)$.

In case of pure states, that is, $V$ is one-dimensional,
Observation~\ref{ob:rel} states that a necessary condition for a pure
state $\ket{\psi}$ to be a unique ground state of some $\K$-local
Hamiltonian is that $\ket{\psi}$ is uniquely determined by its
$k$-RDMs of particles in all $K_j\in\cal{K}$.

As a simple example, consider the one-dimensional space $V$ which is spanned by the
three-qubit $GHZ$ state, given by Eq.(\ref{eq:GHZ}).  Because
$\rho_{V}$ is not $\K$-correlated as discussed in a previous example,
there does not exist a $\K$-local Hamiltonian whose unique ground state is
$\ket{GHZ}$.

Observation~\ref{ob:rel} tells us that in order to find the desired $\K$-local 
Hamiltonian for a given space $V$, first of all $V$ must be $\K$-correlated.
Therefore a $\K$-correlated space is then a natural starting point for talking
about the general problem of `from ground states to local Hamiltonians'. 

One would then wonder whether the necessary condition of
$\K$-correlatedness for a space $V$ being a ground-state space of some
$\K$-local Hamiltonian is also sufficient, which indeed gives rise to
the main question we will discuss in this paper, that we highlight below.
\vspace{.3cm}

\noindent\textbf{Main Question}: \textit{Given a $\K$-correlated space $V$,
does there exist a $\K$-local Hamiltonian which has $V$ as its ground-state space, and
if yes, how can we construct such a Hamiltonian?}

\vspace{.3cm}
%We now discuss some examples to demonstrate this relationship
%between $\cal{K}$-correlated spaces and $\K$-local Hamiltonians given
%by Observation~\ref{ob:rel}.

%For $\K=\{\{1,2\},\{2,3\}\}$, consider a $\K$-local Hamiltonian
%\begin{equation}
%  H_{\K} = - Z_1 Z_2 - Z_2 Z_3, 
%\end{equation}
%The ground-space $V$ of $H_{\K}$ is spanned by
%$\{\ket{000},\ket{111}\}$, so $V$ is $\K$-correlated.  On the other
%hand, as discussed in a previous example, because the maximally mixed
%state $\rho_c$ supported on $V$, given by Eq.(\ref{eq:rhoc}), is
%$\K$-correlated, we can also conclude that $V$ is $\K$-correlated.

%\subsection{$\K$-correlated spaces and convex geometry}

%Observation~\ref{ob:rel} tells us that in order to find the desired $\K$-local 
%Hamiltonian for a given space $V$, first of all $V$ must be $\K$-correlated.
%Therefore a $\K$-correlated space is then a natural starting point for talking
%about the general problem of `from ground states to local Hamiltonians'. 

%One would then wonder whether the necessary condition of
%$\K$-correlatedness for a space $V$ being a ground-state space of some
%$\K$-local Hamiltonian is also sufficient. Unfortunately, in general this might not
%be true. This makes the `from ground states to local Hamiltonians'
%problem rather difficult, even conceptually. 

Unfortunately, this question seems difficult to answer in general. 
In seeking for a better understanding, we start from examing a nice property of 
$\K$-correlated spaces, given by the following observation.
\begin{obs}
\label{obs:kcorr}
For a $\K$-correlated space $V$ and any state $\rho$ supported on $V$, any
state $\sigma$ in $A_{\K}(\rho)$ is also supported on $V$.
\end{obs}

To see why it is the case, denote the range of $\rho$ by $\range(\rho)$ , 
which is the space spanned by all
the eigenstates of $\rho$ with non-zero eigenvalues.  Since $V$ is
$\K$-correlated, we know that the maximally mixed state $\rho_V$
supported on $V$ satisfies $\tilde{\rho}_{V,\K}=\rho_V$.  Therefore,
$\range({\hat{\rho}})\subseteq\range({\rho_V})$ for any $\hat{\rho}\in
A_{\K}(\rho_V)$.  Now for any $\rho$ supported on $V$, we have
$\range({\rho})\subseteq\range({\rho_V})$. Consequently, for any
$\sigma\in A_{\K}(\rho)$, we have
$\range(\sigma)\subseteq\range({\rho_V})$, meaning that $\sigma$ is
also supported on $V$. Note that for this argument there are indeed
some subtle points need to be clarified.  We then include a complete
proof of this observation in Appendix.

%Note that despite the statement of Observation \ref{obs:kcorr} seems to be 
%related to the statement of Observations \ref{obs:ground} and \ref{ob:rel} (indeed 
%Observations  \ref{ob:rel} and \ref{obs:kcorr}
%can give the result in Observation \ref{obs:ground}), the validity
%of \ref{obs:kcorr} does not guarantee that a $\K$-correlated 
%space $V$ to be the ground-state space of some $\K$-local Hamiltonian.
%This may seem not that obvious at the first glance.
%In order to understand where the problem might possibly
%happen, we now introduce a geometric view point for $\K$-correlated
%spaces (note that these kind geometric viewpoints for quantum state
%spaces have been discussed in other literatures, for instance,
%~\cite{VC06,FNW92}). 

Next, we build a connection between $\K$-correlated spaces and
faces of the convex set $\D_{\K}$, which is given by the following
observation.
\begin{obs}
  \label{ob:face}
  For a $\K$-correlated space $V$, $F_{V}$ is a face of the convex set
  $\D_{\K}$.
\end{obs}

To show that this observation holds, first note that it is obvious
that $F_{V}$ is a convex set.  Then for two states $\rho_0$ and
$\rho_1$, let $L$ be a line segment in $\D_{\K}$ with end points
$\vec{R}_{\K}(\rho_0)$ and $\vec{R}_{\K}(\rho_1)$. If $L$ intersects
$F_{V}$ at a point $(1-p)\vec{R}_{\K}(\rho_0) + p\vec{R}_{\K}(\rho_1)$
for some $p\in(0,1)$, then $\exists \sigma$ supported on $V$ such that
$(1-p) \rho_0 + p \rho_1 \in A_\K (\sigma) $. When $V$ is
$\K$-correlated, we have
\begin{equation*}
  \range\bigl((1-p)\rho_0 + p\rho_1\bigr) \subseteq V,
\end{equation*}
and therefore both $\range(\rho_j)$s are spaces of $V$. It then
follows that the entire line segment $L$ is in $F_{V}$.

Note that it is straightforward to show that the reverse of this
observation is also true. That is, for any face $F_{V}$ of $\D_{\K}$,
$V$ is $\K$-correlated.

Observation~\ref{ob:face} characterizes the image $F_{V}$ in $\D_\K$ of a
$\K$-correlated space $V$ as a face of the convex set $\D_{\K}$.  And we
know that ground-state spaces of $\K$-local
Hamiltonians correspond to \textit{exposed} faces of $\D_{\K}$. 
Therefore, the question of whether a $\K$-local Hamiltonian exists to have 
the given $\K$-correlated space as its ground-state space 
then becomes to determine whether the corresponding
face $F_{V}$ is exposed in $\D_{\K}$. 
We examine this question further in the next section.

\section{Non-exposed faces}
\label{sec:nonexpose}

We know that for a general convex set $C$, there does exist non-exposed faces. 
An example is shown in Fig. 1.
%\begin{figure}[h]
%\label{fig:example}
%\includegraphics[scale=1.2]{fig-1.eps}
%\caption{The convex set is the union of a half disk on the right and a
%  rectangle on the left. Point $A$ is by definition a face of this
%  convex set, yet there is no line that touches the convex set only at
%  $A$.}
%\end{figure}
However, for a given interaction pattern $\K$, the geometry of
$\D_\K$ is in general difficult to analyze. Indeed we know that for
local Hamiltonian problems of practical interests, even with the existence 
of a quantum computer, the membership of $\D_\K$ is very difficult to determine
~\cite{Liu06}.

Here we just try to get a bit further to analyze an artificial example. 
We consider a two-qubit system.
In this case, in stead of only requiring that we want a $\K$-local Hamiltonian,
we further want a $\K$-local Hamiltonian of certain type. More precisely,
we want a one-body Hamiltonian $H$ which can only have local terms of $H_1$
and $H_2$ as given below.
\begin{eqnarray}
H_1&=&X_2+\frac{1}{2}(I+Z_1),\nonumber\\
H_2&=&Y_2.
\end{eqnarray}

Now for any given two-qubit state $\ket{\psi}$ which can be uniquely determined
by its mean values on $H_1$ and $H_2$, we wonder whether there exists a
Hamiltonian $H_{\psi}=\alpha H_1+\beta H_2$ that has $\ket{\psi}$ as its unique
ground state. Note that in this case, such a $\ket{\psi}$ is a natural analog of
a $\K$-correlated state and it corresponds to an extreme point of the two-dimensional
convex set given by all points of $(x=\Tr(\rho H_1),y=\Tr(\rho H_2))$, where
$\rho$ is any two-qubit quantum state. This convex set is shown in Fig. 1.
\begin{figure}[htbp]
\includegraphics[scale=2.0]{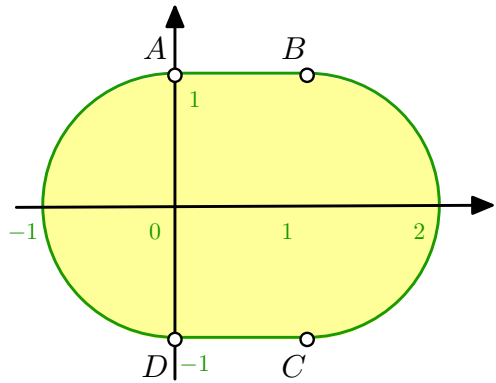}
\caption{The convex set of points given by $(x=\Tr(\rho H_1),y=\Tr(\rho H_2))$.
The convex set is the union of two half disks on the left and right and a
rectangle in the middle. Points $A,B,C,D$ are by definition faces of this
convex set, yet for each point there is no line that touches the convex set only at
the point.}
\label{fig:example}
\end{figure}

What we can see from Fig. 1 is that the there are four non-exposed extreme points $A,B,C,D$ (with coordinates $(0,1),\ (1,1),\ (1,-1),\ (0,-1)$). If we denote $\ket{0_y},\ket{1_y}$ the eigenstates of $Y$ with eigenvalues $\pm 1$ respectively, then these four non-exposed extreme points correspond to quantum states $\ket{0}\ket{0_y},\ket{1}\ket{0_y}, \ket{1}\ket{1_y},\ket{0}\ket{1_y}$, respectively. For each of these four states, apparently it cannot be unique ground state of any kind of Hamiltonian with terms of $H_1$ and $H_2$ only, as $\ket{0}\ket{0_y}$ must be always degenerate with $\ket{1}\ket{0_y}$, and $\ket{0}\ket{1_y}$ must be degenerate with $\ket{1}\ket{1_y}$.

This example is somewhat artificial as one can certainly find a one-body Hamiltonian
which has, for instance, $\ket{0}\ket{0_y}$ as its unique ground if we do not restrict on the terms
of $H_1$ and $H_2$ only. However, it is unclear whether such a relaxation to 
allow any $\K$-local terms is enough to remove all non-exposed faces in general.
Either yes or no would require more deep physical insight beyond a general geometric
analysis of these restricted kind of Hamiltonians. On the other hand, in practice there
might also be physical situations which restricts the form of the terms appearing in an
$\K$-local Hamiltonian (e.g. symmetry restrictions), where a non-exposed face
situation might possibly arise.

% For the convex set $\D_{\K}$, the possibility of existence of a face
% similar to point $A$ in Fig. 1 indicates that for a given
% $\K$-correlated space $V$, there might not exist a $\K$-local
% Hamiltonian $H$ such that $V$ is the exact ground-state space of $V$.
% However, so far we do not know any example to show this could really
% happen.  Therefore it remains open whether the necessary condition of
% $\K$-correlatedness for a space $V$ being a ground-state space of some
% $\K$-local Hamiltonian is also sufficient.

%We would like to emphasize that this geometric view point of
%$\K$-correlated spaces is useful in practical situations. Here we
%discuss some applications. 

In practice, for a given $\K$-correlated space $V$, 
we may circumvent the ``existence analysis'' and anyway go ahead
trying to construct the corresponding $\K$-local Hamiltonian. 
The geometric view point of exposed/non-exposed faces does give
some clue on how to do that. We then discuss a method of perturbation
of finding a $\K$-local Hamiltonian $H$ for a given $\K$-correlated
space $V$, based on this geometric point of view, 
in case there indeed exists such an $H$.

An illustration of the idea is given in Fig. 2. For a given
$\K$-correlated space $V$, our goal is to find some $\K$-local
Hamiltonian $H$ such that the ground space of $H$ is exactly $V$. As
we have already mentioned, this is equivalent to finding some point
$\vec{H}$ in $\P_{\K}$ such that $H\bigcap \D_{\K} = F_V$ where $H$ is
the hyperplane defined by $\vec{H}$.
\begin{figure}[htbp]
\label{fig:idea}
\includegraphics[scale=1.2]{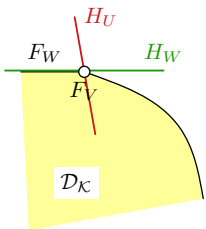}
\caption{$H_W$ is a line that touches
    the convex set $\D_{\K}$ at the top line segment $F_W$ while $H_U$ goes
    through point $F_V$, but doesn't contain points in $F_W\setminus F_V$.}
\end{figure}
As $F_V$ is a face of $\D_{\K}$, we can find a hyperplane that
contains $F_V$. Let this hyperplane be $H_W$ and a corresponding point
in $\P_{\K}$ be $\vec{H}_W$ where $W$ is the ground space of the
corresponding Hamiltonian $H_W$. We know that $W$ is also
$\K$-correlated as $H_W$ is a $\K$-local Hamiltonian. If $W$ equals
$V$, we are done. Otherwise, one sees that the intersection of
hyperplane $H_W$ and $\D_{\K}$ is exactly the face $F_W$. Moreover,
$F_V$ is a face of $F_W$.

Now we wish to find a perturbation $\K$-local Hamiltonian $H_U$ which
can `split' the energy of states supported on $V$ and those supported
on $W\setminus V$, where the Hamiltonian $H=t H_W + H_U$ can have $V$
as its exact ground-state space for large enough $t$. We show that the
following conditions for $H_U$ is sufficient.
\begin{enumerate}
\item $F_V$ is in the kernel of $H_U$
\item $\ip{\vec{R}_{\K}(\rho)}{\vec{H}_U}>0$ for all
  $\vec{R}_{\K}(\rho) \in F_W\setminus F_V$.
\end{enumerate}

Once there exists an $H_U$ satisfies these two conditions, we can show
that the Hamiltonian $H=t H_W + H_U$ can have $V$ as its exact
ground-state space for large enough $t$. Let $U$ be the kernel of
$H_U$.  Denote $\lambda,\mu$ as the smallest positive eigenvalue of
$H_W$ and $H_U$ respectively, and $\omega = \|H_U\|$ as the operator
norm of $H_U$. For any $n$ particle state $\ket{\psi}$, write it as
$\sum_{j=1}^3 \ket{\psi_j}$ such that $\ket{\psi_1}\in U\cap W$,
$\ket{\psi_2} \in W\cap U^\perp$, and $\ket{\psi_3}\in
W^\perp$. Therefore, we have
\begin{equation*}
  \begin{split}
    & \bra{\psi}(tH_W+H_U)\ket{\psi}\\
    =\; & t\bra{\psi_3}H_W\ket{\psi_3} + (\bra{\psi_2}+\bra{\psi_3})
    H_U
    (\ket{\psi_2}+\ket{\psi_3})\\
    \ge\; & (t\lambda-\omega) \|\psi_3\|^2 - 2\omega \|\psi_2\|
    \|\psi_3\| + \mu \|\psi_2\|^2.
  \end{split}
\end{equation*}
The above equation is non-negative if $\|\psi_3\|=0$. Otherwise, it is
a quadratic function and is positive for large enough $t$. This means
that the ground space of $tH_W+H_U$ is $W\cap U=V$.

In general, we do not know whether such an $H_U$ does exist. 
However, as we will show in Sec.~\ref{sec:FF}, in practical situations this method
of perturbation can indeed help us find a $\K$-local $H$ for a
given $\K$-correlated space $V$, in case there indeed exists such an
$H$. On the other hand, if one fails to find such an $H_U$,
then this indicates the existence of a non-exposed face.

Finally, we mention another direct application of the geometric
viewpoint of $\K$-correlated spaces, given by the following
observation.
\begin{obs}
  \label{ob:cap}
  The nontrivial intersection of two $\K$-correlated spaces $V_1$ and
  $V_2$ is $\K$-correlated. Furthermore, for two spaces $V_1$ and
  $V_2$ which are ground-state spaces of the $\K$-local Hamiltonians
  $H_1$ and $H_2$ respectively, then the nontrivial intersection of
  $V_1$ and $V_2$ is a ground space of some other $\K$-local
  Hamiltonian.
\end{obs}
To show why this is true, simply note that the intersection of two
faces $F_{V_1}$ and $F_{V_2}$ of $\D_{\K}$ is another face $F_V$,
where $V=V_1\cap V_2$. And because $F_V$ is the intersection of two
faces, one can use the method of perturbation to find the $\K$-local
Hamiltonian for $F_V$, where the existence of $H_U$ is ready to
verify. Indeed, the Hamiltonian which has $V$ as its exact
ground-state space can be simply chosen as $H_1+H_2$.

\section{From frustration-free to frustrated systems}
\label{sec:FF}

In this section, we discuss a method of finding a corresponding
$\K$-local Hamiltonian for some given $\cal{K}$-correlated spaces,
which is related to local frustration-free Hamiltonians.  This will
allow us to find the desired Hamiltonians for some spaces whose
correlation patterns are well known, including the three-qubit pure
states (the topic of subsection A), and the $n$-qubit $W$-type states
(the topic of subsection B).

Given a space $V$ and a pattern $\K$, let $\rho_V$ be the maximally
mixed state of $V$, and denote $\rho_V({K_i})$ the $k$-RDMs of
$\rho_V$ for particles in $K_i\in\K$.  Let
$V_{K_i}=\range(\rho_V({K_i}))$.  If
\begin{equation}
  V=\bigcap\limits_{K_i\in\K}V_{K_{i}}\otimes I_{\bar{K}_{i}},
\label{eq:FF}
\end{equation}
where $\bar{K}_{i}=\{1,2,\ldots,n\}\setminus{K}_{i}$, then $V$ is the
ground-state space of the $\K$-local Hamiltonian
\begin{equation}
  H=\sum\limits_{K_i\in\K} H_{K_i},
\end{equation}
where $H_{K_i}$ is the projection onto the kernel of $\rho_V({K_i})$.

The Hamiltonian $H$ is known to be frustration-free, as the
ground-state space $V$ of $H$ is also the ground-state space of each
term in the summation. That is, $V_{K_{i}}$ is the ground-state space
of $H_{K_i}$. We call this kind of $\K$-local Hamitonian
$\K$-frustration-free ($\K$-FF). By Observation~\ref{ob:rel}, we know
that the ground-state space $V_{\K}$ of a $\K$-FF Hamiltonian must be
$\K$-correlated.

Although in general whether a space $V$ is the ground-state space of a
$\K$-FF Hamiltonian could be difficult to analyze, that is,
Eq.(\ref{eq:FF}) is not easy to check for large systems, at least in
principle this provides a way of characterizing these kind of spaces
as well as finding the corresponding $\K$-FF Hamiltonians.  This can
then be used as a starting point to find a $\K$-local Hamiltonian for
a space $V'\subset V$ that is known to be $\K$-correlated, using the
perturbation method discussed in Sec.~\ref{sec:rel}.  The idea is, we
choose the unperturbed Hamiltonian $H_0$ as the $\K$-FF Hamiltonian
$H$, then we will need to find a $\K$-local Hamiltonian $H_{1}$ such
that the null space of $H_{1}$ contains $V'$ and for any state
$|\phi\rangle\in V-V'$, $\langle\phi|H_{1}|\phi\rangle>0$.  Then for
large enough $t$, the $\K$-local Hamiltonian $tH_0+H_{1}$ will have
$V'$ as its ground-state space.

To demonstrate the application of these methods, we consider two
examples.  Our first example is the case of three qubits that we will
discuss in subsection A. And the second example are those $W$-type
states that we will discuss in subsection B.

\subsection{The three-qubit case}
\label{sec:threequbit}

It is well-known that almost all three-qubit pure states are uniquely
determined by their $2$-RDMs except those states which are local
unitary (LU) equivalent to $GHZ$-type states
$\alpha\ket{000}+\beta\ket{111}$~\cite{LPW02}.  That is, almost all
three-qubit pure states are $\K'$-correlated for
\begin{equation}
  \K'=\{\{1,2\},\{2,3\},\{1,3\}\}.
\end{equation}

Here we will then find the $\K'$-local Hamiltonian ${H}$ for all
three-qubit states, starting from a $\K'$-FF Hamiltonian and using the
perturbation method, except for those states which are LU equivalent
to $GHZ$-type states. Indeed, our method finds some $\K$-local
Hamiltonians for these states, where
\begin{equation}
  \K=\{\{1,2\},\{2,3\}\}\subset\K'.
\end{equation}
This means that indeed all three-qubit pure states are uniquely
determined by their $2$-RDMs of particles $\{1,2\}$ and $\{2,3\}$,
except for those states which are LU equivalent to $GHZ$-type
states. In other words, only two out of the three $2$-RDMs are enough
to uniquely determine these states, which is an improvement of the
results given in~\cite{LPW02}.

Note that one of the standard forms for a three-qubit pure state up to
LU transformation is \cite{Acin}
\begin{equation}
  \ket{\psi}_{123} = \lambda_{0} \ket{000} + \lambda_{1} \ket{100} +
  \lambda_{2} \ket{101} + \lambda_{3} \ket{110} + \lambda_{4}
  \ket{111}
  \label{eq:3qubit}
\end{equation}
where $\lambda_{1}$ is complex, and $ \lambda_{0}, \lambda_{2},
\lambda_{3}, \lambda_{4} $ are real.

We start from constructing a $\K$-FF Hamiltonian $H_{\K}$ for $\K =
\{\{1,2\}, \{2,3\}\}$ which contains $\ket{\psi}_{123}$ as a ground
state. Define the space $\mathcal{S}$ as
\begin{eqnarray}
  \mathcal{S} = (\range( \gamma_{\{1,2\}} ) \otimes I_{\{3\}}) & \cap
  & (I_{\{1\}} \otimes \range( \gamma_{\{2,3\}} )),
\end{eqnarray}
 where $ \gamma_{\{i,j\}} $ is the $2$-RDM of $\ket{\psi}_{123}$
of particles $\{i,j\}$.

It is straightforward to show that $\mathcal{S}$ is always
two-dimensional for any entangled $\ket{\psi}_{123}$. That is,
$\ket{\psi}_{123}$ cannot be written as a product of a single and a
two-qubit state.  In this case, one always has
\begin{equation}
  \ket{\psi'}_{123} = \ket{1} \otimes (\lambda_{2} \ket{0} +
  \lambda_{4} \ket{1}) \otimes (\lambda_{3} \ket{0} + \lambda_{4}
  \ket{1}) \in \mathcal{S} 
  \label{eq:prod}
\end{equation}
That is, $\mathcal{S}$ always contains a product state
$\ket{\psi'}_{123}$ (see~\cite{CCD+10}).

We can then choose a $\K$-FF Hamiltonian
\begin{equation} 
  H_0=H_{\{1,2\}}+H_{\{2,3\}},
\end{equation}
where $H_{\{i,j\}}$ is the projection onto the kernel of
$\gamma_{\{i,j\}}$.  Then $\mathcal{S}$ is the ground-state space of
$H_0$, which is two-dimensional and spanned by $\ket{\psi}_{123}$ and
$\ket{\psi'}_{123}$.

Now we need to find a perturbation $\K'$-local Hamiltonian $H_{1}$
such that for large enough $t$, the $\K'$-local Hamiltonian
$tH_0+H_{1}$ has $\ket{\psi}_{123}$ as its unique ground state. First
define
\begin{eqnarray}
  \ket{\phi}_{123} & = & \lambda_1 \ket{100} + \lambda_2 \ket{101} +
  \lambda_3 \ket{110} + \lambda_4 \ket{111} \nonumber\\
  & = & \ket{1} \otimes ( \lambda_1 \ket{00} + \lambda_2 \ket{01} +
  \lambda_3 \ket{10} + \lambda_4 \ket{11} ) \nonumber\\
  & = & \ket{1} \otimes \ket{\tilde{\psi}}_{23}.
\end{eqnarray}

For the generic case, $\ket{\phi}_{123}$ is linear independent of
$\ket{\psi'}_{123}$, which means
\begin{equation}
  \label{eq:ind}
  \lambda_{1}\lambda_{4}\neq\lambda_{2}\lambda_{3}.
\end{equation}

Now define 
\begin{equation}
  \eta = _{123} \bra{\psi'} \phi \rangle_{123} = \lambda_{1}
  \lambda_{2} \lambda_{3} + \lambda_{2}^{2} \lambda_{4} +
  \lambda_{3}^{2} \lambda_{4} + \lambda_{4}^{3},
\end{equation}
and choose
\begin{equation}
\ket{\xi}_{123} = \ket{\psi'}_{123} - \eta^{*} \ket{\psi}_{123}.
\end{equation}

Note that
\begin{equation}
  _{123}\bra{\psi}\xi\rangle_{123}=0,
\end{equation}
and $\ket{\xi}_{123}$ has a form 
\begin{equation}
  \ket{\xi}_{123} = \alpha \ket{000} + \beta \ket{1} \otimes \ket{\tilde{\xi}}_{23},
\end{equation}
where $\ket{\tilde{\xi}}_{23}$ is a pure state of particles $\{2,3\}$
which is linear independent of $\ket{\tilde{\psi}}_{23}$ according to
Eq.(\ref{eq:ind}).

We can then choose a two-particle Hermitian operator $H'_{\{2,3\}}$
acting on particles $\{2,3\}$ such that $\ket{00}_{23}$ and
$\ket{\tilde{\psi}}_{23}$ span the kernel of of $H'_{\{2,3\}}$, hence
$_{23}\bra{\tilde{\xi}}H\ket{\tilde{\xi}}_{23}=r\neq 0$. So the
perturbation Hamiltonian $H_{1}$ can be just chosen as
$H_{1}=H'_{\{2,3\}}$ if $r>0$ or $H_{1}=-H'_{\{2,3\}}$ if $r<0$. Then
for large enough $t$, $tH_0+H_{1}$ has $\ket{\psi}_{123}$ as its
unique ground state.

For the case $\ket{\phi}_{123}$ is linear dependent of
$\ket{\psi'}_{123}$, which in general means
\begin{equation}
  \label{eq:dep}
  \lambda_{1}\lambda_{4}=\lambda_{2}\lambda_{3},
\end{equation}
we can also find $H_{1}$ in this case, unless $\ket{\phi}_{123}$ is LU
equivalent to the $GHZ$-type state. Note that Eq.(\ref{eq:dep})
indicates that $\lambda_1$ is real.

We can rewrite $\ket{\psi}_{123}$ as 
\begin{equation}
  \ket{\psi}_{123}=\lambda_{0}\ket{000}+\lambda_{xy}\ket{1xy},
\end{equation}
where
\begin{eqnarray}
  \ket{x} & = & x_{0}\ket{0}+x_{1}\ket{1},\nonumber \\
  \ket{y} & = & y_{0}\ket{0}+y_{1}\ket{1},
\end{eqnarray}
with $x_0,x_1,y_0,y_1$ real, $x_{0}^{2}+x_{1}^{2}=1$ and
$y_{0}^{2}+y_{1}^{2}=1$.

We know that $\ket{1xy}$ is also in the ground-state space of $H_0$,
so the ground-state space of $H_0$ of is actually spanned by two
orthogonal product states $\ket{000}$ and $\ket{1xy}$.

In general, when $\ket{\psi}_{123}$ is not LU equivalent to the
$GHZ$-type state, we have
\begin{equation}
\bra{0}x\rangle\neq0,\ \text{or}\ \bra{0}y\rangle\neq0.
\end{equation}
Without loss of generality, we assume $\bra{0}y\rangle\neq0$, that is,
$y_0\neq 0$.

Now we need to find some Hamiltonian ${H}_1$ to `split' $\ket{000}$
and $\ket{1xy}$ such that such that for large enough $t$ the ground
state of $t{H}_0+ H_{1}$ could be uniquely $\ket{\psi_{123}}$, based
on the perturbation method.  We show this is always possible. Let
\[ D_1=
\begin{pmatrix}
  \frac{\lambda_{xy}}{\lambda_0} & 0  \\
  0 & \frac{\lambda_{0}}{\lambda_{xy}}
\end{pmatrix}, M_2=
\begin{pmatrix}
  x_0 & x_1  \\
  x_1 & -x_0
\end{pmatrix},\ M_3=
\begin{pmatrix}
  y_0 & y_1  \\
  y_1 & -y_0
\end{pmatrix},\] 
then we have
\begin{eqnarray}
  X_1D_1\otimes M_2\otimes M_3\ket{\psi_{123}}=\ket{\psi_{123}},
\end{eqnarray}
which gives
\begin{equation}
  X_1 D_1 \otimes  M_2 \otimes I_3 \ket{\psi_{123}} = I_1 \otimes I_2
  \otimes M_3 \ket{\psi_{123}},
\end{equation}
where $I_j$ is the identity operator acting on the $j$th particle.

Now we can choose a two-particle operator 
\begin{equation}
  H'_1=X_1D_1\otimes M_2\otimes I_3-I_1\otimes I_2\otimes M_3,
\end{equation}
then $\ket{\psi_{123}}$ is in the kernal of $H'_1$ and
$\bra{000}H'_1\ket{000}\neq 0$.

Let $ \ket{\psi^{\perp}}_{123} = \lambda_{xy} \ket{000} - \lambda_{0}
\ket{1xy} $, and $_{123} \bra{\psi^{\perp}} H'_1
\ket{\psi^{\perp}}_{123} = r \neq 0 $, then the perturbation
Hamiltonian $ H_{1} $ can be just chosen as $ H_{1} = H'_1 $ if $ r >
0 $ or $ H_{1} = - H'_1 $ if $ r<0 $.  Then for large enough $ t $, $
t H_0 + H_{1} $ has $ \ket{\psi}_{123} $ as its unique ground state.

Similar procedure works if $ \bra{0} y \rangle = 0 $ but $ \bra{0} x
\rangle \neq 0 $. The procedure will fail to result in having $
\ket{\psi}_{123} $ as the unique ground state $ t H_0 + H_{1} $ for
any $ t $ if both $ \bra{0} x \rangle \neq 0 $ and $ \bra{0} y \rangle
\neq 0 $. In that case, one will have $ _{123} \bra{\psi^{\perp}} H'_1
\ket{\psi^{\perp}}_{123} = 0 $. In that case, by properly chosen $ t
$, one can have $ \ket{\psi}_{123} $ as the nondegenerate first
excited state $ t H_0 + H_{1} $ ~\cite{CJW+11}.

To summarize, we have found the $\K$-local Hamiltonian for all
three-qubit pure states for $\K=\{\{1, 2\}, \{2, 3\}\}$, except for those
states which are LU equivalent to $GHZ$-type states.  By
Observaion~\ref{ob:rel}, our result also shows that all three-qubit
pure states are uniquely determined by their $2$-RDMs of particles
$\{1, 2\}$ and $\{2, 3\}$, except for those states which are LU
equivalent to $GHZ$-type states.

\subsection{The $W$-type states}

In this subsection we discuss the $n$-qubit $W$-type states
$\ket{W(n)}_{type}$,
\begin{equation}
  \ket{W(n)}_{type}=\sum_{i=1}^{n}a_s\ket{\mathbf{r}_i},
\end{equation}
where $\mathbf{r}_i$ is the $n$-bit strings with the $i$-th coordinate
$1$ and all the other coordinates $0$, $a_i\neq 0$ and
$\sum_{i=1}^n|a_i|^2=1$.

It is known that $\ket{W(n)}_{type}$ is uniquely determine by its
$2$-RDM~\cite{PR08}.  What is more, any $n-1$ out of the ${n\choose
  2}$ $2$-RDMs are sufficient to uniquely determine
$\ket{W(n)}_{type}$, so we can actually put the $n$-qubit on a
one-dimensional chain and consider only the $2$-RDMs of all the
nearest neighbour pairs.  More precisely, let
\begin{equation}
  \K=\{\{1,2\},\dots,\{n-1,n\}\},
\end{equation}
then the $W$-type states are  $\K$-correlated.

Here we discuss how to find the $\K$-local Hamiltonian whose unique
ground state is a given $W$-type state. We start from the three-qubit
case. In Sec.~\ref{sec:threequbit}, we have already solved this
problem for all three-qubit pure states. Here we re-examine the $W$
state case so we understand how to generalize it to the general
$n$-qubit case. We start from the fact that the three-qubit $W$-type
state can be written as
\begin{equation}
  \ket{W(3)}_{type}=a_{1}\ket{001}+a_{2}\ket{010}+a_{3}\ket{100},
\end{equation}
 and observe that 
\begin{eqnarray}
  (I_{\{1\}}\otimes\range(\gamma_{\{2,3\}})) & \cap & (\range(\gamma_{\{1,2\}})\otimes I_{\{3\}})\nonumber \\
  & = & \text{span}\{\ket{W}_{type},\ket{000}\},
\end{eqnarray}
 and 
\begin{equation}
  \bra{000}W(3)\rangle_{type}=0.
\end{equation}

We can first choose a $\K$-FF Hamiltonian
\begin{equation}
  \label{eq:KFFW3}
  {H}_0=H_{\{1,2\}}+H_{\{2,3\}}
\end{equation}
for $\K=\{\{1,2\},\{2,3\}\}$. Here $H_{\{i,j\}}$ is the projection
onto the kernel of $\gamma_{\{i,j\}}$.  Then
$\{\ket{W(3)}_{t},\ket{000}\}$ spans the two-dimensional ground-state
space of $H_0$.

We can then choose
\begin{equation} 
  H_{1}=-Z_{1}-Z_{2}-Z_{3}.
\end{equation}
For a large enough $t$, we have $\ket{W(3)}_{type}$ is the unique
ground state of $tH_0+H_{1}$.

Now we take a look at the special case where $a_{1}=a_{2}=a_{3}$, so
$\ket{W(3)}_{type}$ becomes the three-qubit $W$-state $\ket{W(3)}$,
where
\begin{equation}
  \ket{W(3)}=\frac {1} {\sqrt{3}} \left( \ket{001} + \ket{010} +
    \ket{100} \right).
\end{equation}
Now the $\K$-FF Hamiltonian $H_0$ given in Eq.(\ref{eq:KFFW3}) has a
two-dimensional ground-state space spanned by
$\{\ket{W(3)},\ket{000}\}$.

Note that now both $H_{\{1,2\}}$ and $H_{\{2,3\}}$ are projections
onto the space spanned by
\begin{eqnarray}
  \ket{\alpha} & = & \ket{11}\nonumber \\
  \ket{\beta} & = & \frac{1}{\sqrt{2}}(\ket{01}-\ket{10}),
\end{eqnarray}
thus $ H_{\{1,2\}} $ can be written as
\begin{equation}
  H_{\{1,2\}} = p_{\alpha} \ket{\alpha} \bra{\alpha} + p_{\beta}
  \ket{\beta} \bra{\beta},
\end{equation}
where $ p_{\alpha}, p_{\beta} > 0 $.

In terms of Pauli operators, $ H_{\{1,2\}} $ has a form 
\begin{eqnarray}
  H_{\{1,2\}} = & - & p_{\alpha} ( X_{1} X_{2} + Y_{1} Y_{2} )\nonumber \\
  & + & ( p_{\beta} - p_{\alpha} ) Z_{1} Z_{2} - p_{\beta} ( Z_{1} +
  Z_{2} ).
\end{eqnarray}
And a similar form holds for $H_{\{2,3\}}$.

This form of $H_{\{i,j\}}$ can be generalized to $n$-qubit case.
To see this, note that
\begin{equation}
\bigcap_{i}\range(\gamma_{\{i,i+1\}})\otimes I_{\{\overline{i,i+1}\}}=\{\ket{W(n)},\ket{00...0}\}.
\end{equation}
 Now we can choose $H_{0}=\sum_{i}H_{\{i,i+1\}}$, 
\begin{eqnarray}
H_{\{i,i+1\}}= & - & p_{\alpha}(X_{i}X_{i+1}+Y_{i}Y_{i+1})\nonumber \\
 & + & (p_{\beta}-p_{\alpha})Z_{i}Z_{i+1}-p_{\beta}(Z_{i}+Z_{i+1}).
\end{eqnarray}
 and $H_{1}=-\sum_{i}Z_{i}$. Then for a large enough $t$, the
$\K$-local Hamiltonian $H_{\K}=tH_{0}+H_{1}$ has the $n$-qubit
$W$-state $\ket{W(n)}$ as its unique ground state.

If we take a periodic boundary condition instead of a chain, that
is, choose
\begin{equation}
\K'=\{\{1,2\},\{2,3\},\dots,\{n-1,n\},\{n,1\}\}, 
\end{equation}
then for a small enough $\epsilon$, the $\K'$-local Hamiltonian
${H}_{w}$ that $\ket{W(n)}$ is a unique ground state of can be written
as 
\begin{eqnarray} 
  H_{w} & = & -\sum_{i,i+1} \left( p_{\alpha} X_{i}
    X_{i+1} + p_{\alpha} Y_{i} Y_{i+1} + ( p_{\alpha} - p_{\beta} )
    Z_{i} Z_{i+1} \right) \nonumber \\
  & - & \sum_{i} (2\beta-\epsilon) Z_{i}.
  \label{eq:XXZ}
\end{eqnarray}

Actually, $\mathcal{H}_{w}$ is of very nice physical meaning as it
is a famous spin model called `Heisenberg $XXZ$ model', where we also
have a term of external magnetic field, which is given by the second sum term
in $\mathcal{H}_{w}$. This model is extensively studied in the literature,
for instance, see \cite{XXZ} and references therein.

Note that our results are consistent with those obtained in \cite{WXX},
where a special case $p_{\alpha}=p_{\beta}$ is considered, so $\mathcal{H}_{w}$
is reduced to a Heisenberg $XX$ chain in a transversal magnetic field.
We observe that although for different values of $p_{\alpha}$ and
$p_{\beta}$, the ground state could be all uniquely $\ket{W(n)}$, the
Hamiltonian $\mathcal{H}_{w}$ do have different spectrums, hence are
different Hamiltonians.

\section{Hamiltonians from thermodynamical ensembles}
\label{sec:TE}

In this section, we discuss a general method to determine whether a
given state space $V$ is $\K$-correlated, if so, we find the
$\K$-local Hamiltonian such that its ground-state space is $V$.  Our
approach is based on the viewpoint of thermodynamical ensembles.

For a given space  $V$, we introduce
\begin{equation}
\rho(p) = p \frac {I} {D} + (1-p) \rho_V,\label{eq:rhop}
\end{equation}
where $I$ is the identity operator acting on the Hilbert space
$\cal{H}$ of the $n$-pariticle system with a finite dimension $D$, and
$\rho_V$ is the maximally mixed state of $V$.  Obviously,
$\rho(0)=\rho_{V}$.

As the state $\rho(p)$ is of full rank for $p\in[1,0)$,
$\trho_{\K}(p)$, as given by Eq.(\ref{eq:tilde}), can be written in an
exponential form~\cite{Zho08}
\begin{equation}
\trho_{\K}(p) = \frac {\exp(-\tilde{H}_\K(p))}
{\mathrm{Tr}\exp(-\tilde{H}_\K(p))},
\label{eq:exp}
\end{equation}
where $ \vec{R}_{\K} ( \trho_{\K}(p) ) = \vec{R}_{\K} ( \rho (p) )$,
and the Hermitian operator $ \tilde{H}_\K ( p ) $ is $ \K $-local. And
indeed such an exponential form is unique~\cite{Zho09a}.

The key observation here is that $\trho_{\K}(p)$ can be viewed as a
thermal equilibrium state corresponding to the $\K$-local Hamiltonian
$H_\K(p)$: we can define $\beta(p)H_{\K}(p)=\tilde{H}_{\K}(p)$ with
$\beta(p)$ a positive constant inversely proportional to temperature.

Note that the maximally mixed state $\rho_V$ of $V$ is an equal weight
mixture of orthonormal pure states, which span $V$.  As $\rho_V$ is
$\K$-correlated, we have
\begin{equation}
  \lim_{p\rightarrow 0} \trho_{\K} ( p ) = \rho_V,
  \label{eq:rhoHp0}
\end{equation}
according to the continuity principle given in~ \cite{Zho08,Zho09a}.
Eq. (\ref{eq:rhoHp0}) then implies that $\rho_V$ is the equal weight
mixture of the ground states of $H_\K(0)$, and the corresponding
temperature goes to $0$, i.e., $\lim_{p\rightarrow0}\beta(p)=+\infty$.

Note that the continuity principle discussed in~ \cite{Zho08,Zho09a}
is an argument, not a rigorous proof. And this method definitely fails
for those $\K$-correlated spaces which is similar to point $A$ in
Fig. 1. However, this viewpoint of thermal equilibrium ensemble gives
a good physical intuition to understand Observation~\ref{ob:rel}.

One numerical method to find $\tilde{H}_{\K}(p)$ for $p\in[1,0)$ can
be developed based on the discussion in~\cite{Zho09b}. The idea is
that if the continuity principle is valid, then when $p$ is
arbitrarily close to one, the ground-state space of
$\tilde{H}_{\K}(p)$ will be also arbitrarily close to $V$.

As an example to test our numerical method, consider the following
$4$-qubit state
\begin{equation}
  \ket{\psi_{1}} = \frac {1} {2} ( \ket{0000} + \ket{0101} + \ket{1000}
  + \ket{1110}).
  \label{eq:psi1}
\end{equation}

Our numerical method shows that there exists a Hamiltonian containing
only one and two particle interaction terms, such that
$\ket{\psi_{1}}$ is the unique ground state. This Hamiltonian can be
given by $p=0.0001$, that is,
\begin{eqnarray*}
  \tilde{H}(0.0001) & = & -3.2390Z_{4}+4.2001X_{3}X_{4}+4.2001Y_{3}Y_{4}\\
  & - & 3.2390Z_{3}-0.5912Z_{3}Z_{4}-6.4827X_{2}X_{4}\\
  & - & 6.4827X_{2}X_{3}+6.4827Y_{2}Y_{4}+6.4827Y_{2}Y_{3}\\
  & + & 6.7571Z_{2}+1.5227Z_{2}Z_{4}+1.5227Z_{2}Z_{3}\\
  & - & 4.2950X_{1}-2.4012X_{1}Z_{4}-2.4012X_{1}Z_{3}\\
  & - & 8.8603X_{1}Z_{2}+4.5280Z_{1}Z_{4}-4.5280Z_{1}Z_{3},
\end{eqnarray*}
and one can readily check $\ket{\psi_{1}}$ is the unique
ground state of $\tilde{H}(0.0001)$.  By Observation~\ref{ob:rel},
$\ket{\psi_1}$ is then $\K(\psi_1)$-correlated for
\begin{eqnarray}
\label{eq:psi}
\K(\psi_1)=\{\{1,2\},\{2,3\},\{1,3\},\{3,4\},\{2,4\},\{1,4\}\}.
\end{eqnarray}

This method also allows us to determine whether a given space $V$ is
$\K$-correlated or not. If the method returns a $\K$-local Hamiltonian
$H(p)$ with $p$ sufficiently small, whose ground-state space is larger
than $V$, then $V$ is not $\K$-correlated. Otherwise it returns
exactly $V$.

As an example, consider the following state 
\begin{equation}
  \ket{\psi_{2}} = \frac {1} {2} ( \ket{0000} + \ket{1011} + \ket{1101} +
  \ket{1110} ).
\end{equation}

Our numerical method shows that there does not exist a Hamiltonian
containing only one and two-particle interaction terms, such that
$\ket{\psi_{2}}$ is the unique ground state. Indeed, the state
\begin{equation}
  \ket{\psi'_{2}} = \frac {1} {2} ( - \ket{0000} + \ket{1011} +
  \ket{1101} + \ket{1110} )
\end{equation}
has the same $\K$-projection as that of $\ket{\psi_{2}}$.
 
One would expect that our numerical method cannot be efficient in
general. Indeed, even in practice, the Hamiltonians that we are
interested in mainly involve only one and two-particle interaction
terms associated with certain lattice geometry, the complexity of our
numerical method grows super exponential with the system size
$n$. Therefore, for each special case considered, one usually needs to
combine this method with some other techniques.

Here we introduce a method of subsystems to reduce the complexity of
the above numerical method for some specific cases, based on the
discussion of frustration-free systems given in Sec.~\ref{sec:FF}.
That is, in some cases, we can start from a $\K$-FF Hamiltonian and
look at the subsystems of each term of the $\K$-FF Hamiltonian.  The
advantage of this method of subsystems is that one can reduce total
dimension of the Hilbert space that one needs to calculate the
$\K$-local Hamiltonians, by using some frustration-free properties of
the quantum space $V$.

Recall that a $\K$-FF Hamiltonian is $\K$-local.  Denote $P(K_i)$ the
power set of $K_i$ for each $K_i\in\K$.  We then define
\begin{equation}
\K'=\bigcup\limits_{i} \K'_i,
\end{equation}
where each $\K'_i$ is a subset of $P(K_i)$.  In practice we will be
interested in some pattern $\K'$ with $|K'_j|=k'$ for $K'_j\in\K'$,
where $k'<k$. In other words, the $\K$-FF Hamiltonian contains
$k$-particle interactions, but the $\K'$-local Hamiltonian we want to
find contains only $k'<k$-particle interactions.

The following observation provides a method of finding a $\K'$-local
Hamiltonian for the ground-state space $V$ of a $\K$-FF Hamiltonian.
\begin{obs}
  \label{ob:sub}
  Given a space $V$ which is the ground-state space of a $\K$-FF
  Hamiltonian.  If for any $K_{i}\in\K$, $\range(\rho_V(K_i))$ is
  $\K'_i$-correlated, then $V$ is $\K'$-correlated.
\end{obs} 
To see how this observation works, for each $K_{i}$,
$\range(\rho_V(K_i))$ is $\K'_{i}$-correlated, so one can find a
$\K'_i$-local Hamiltonian $H_{\K'_{i}}$ which has
$\range(\rho_V(K_i))$ as its ground-state space. However these spaces
of $\range(\rho_V(K_i))$ determines $V$,
i.e. $\bigcap_{i}\range(\rho_V(K_i))=V$, so the Hamiltonian
$\sum_{i}H_{\K'_{i}}$ has $V$ as its ground-state space.

As an example, consider the state $\ket{\psi_{1}}$ given in
Eq.(\ref{eq:psi1}).  It is straightforward to show that
$|\psi_{1}\rangle$ is the unique ground state of a $\K$-FF Hamiltonian
for $\K=\{\{1,2,3\},\{2,3,4\}\}$.  However, this will give us a
non-practical Hamiltonian which involves three-particle interactions.

Note that the space $V_{\{1,2,3\}}=\range(\ket{\psi_1}\bra{\psi_1}(\{1,2,3\}))$ 
is spanned by 
\begin{equation}
  V_{\{1,2,3\}}=\text{span}\{\ket{000}+\ket{110}+\ket{111},\
  \ket{010}\}
  \label{eq:V123}
\end{equation}
and the space $V_{\{2,3,4\}}=\text{ker}(\ket{\psi_1}\bra{\psi_1}(\{2,3,4\}))^{\perp}$ 
is spanned by 
\begin{equation}
  V_{\{2,3,4\}} = \text{span} \{\ket{000}+\ket{101},\
  \ket{000}+\ket{110}\}.
  \label{eq:V234}
\end{equation}

We can now use our numerical method to further show that
$V_{\{1,2,3\}}$ is $\{\{1,2\},\{2,3\},\{1,3\}\}$-correlated, and
$V_{\{2,3,4\}}$ is $ \{ \{2,3\}, \{3,4\}, \{2,4\} \}
$-correlated. Therefore, by Observation~\ref{ob:sub}, $\ket{\psi_1}$
is $\K'(\psi_1)$-correlated for
\begin{eqnarray}
  \label{eq:psiprime}
  \K'(\psi_1)=&&\{\{1,2\},\{2,3\},\{1,3\},\{3,4\},\{2,4\}\}.
\end{eqnarray}

In this example, we use the method of subsystems to reduce the
calculation in our algorithm for a $n=4$ state to two $n=3$
spaces. One could expect for larger system which are ground-state
space of some local frustration-free Hamiltonians involving at most
$k$-particles interactions, this method of subsystems may further
reduce the calculation in our numerical method from a large $n$ to
some small number $k$.  Moreover, recall Eq.(\ref{eq:psi}), we
actually have $\K'(\psi_1)\subset\K(\psi_1)$, so the result obtained
by this method of subsystems gives a slightly simpler interaction
pattern of the Hamiltonian.

Finally, as a remark, note that the reverse of
Observation~\ref{ob:sub} is not true, as the space
$V_{\{1,3,4\}}=\range(\ket{\psi_1}\bra{\psi_1}(\{1,3,4\}))$, spanned
by
\begin{equation}
\label{eq:V134}
V_{\{1,3,4\}}=\text{span}\{\ket{000}+\ket{100},\ \ket{001}+\ket{110}\}
\end{equation}
is not  $\{\{1,2\},\{3,4\},\{1,4\}\}$-correlated.

\section{Conclusion and Discussion}
\label{sec:CD}

In this paper, we raised an interesting questions of ``from ground states to local Hamiltonians''.
That is, for a given space $V$, one would like to know whether $V$ can be the ground-state space
of some $k$-local Hamiltonian $H=\sum_i H_i$ which contains only
certain terms of $k$-particle interactions; and if such a $k$-local
Hamiltonian exists, how to find it. 
As a starting point, it turns out that a natural necessary condition for a space
$V$ to be the ground-state space of some local Hamiltonian with a
given interaction pattern, is that the maximally mixed state supported
on $V$ is uniquely determined by its reduced density matrices
associated with the given pattern, based on the principle of maximum
entropy. This simple observation builds an interesting link between correlations of quantum
states and ground-state space of local Hamiltonians. 
%Unfortunately,
%it is unclear whether this necessary condition is in general also sufficient.

We have introduced the concept of $\K$-correlated spaces 
and explained its physical and geometric meaning.  
By introducing the concept of $\K$-local Hamiltonians which
describe local Hamiltonians with given interaction patterns in a more
formal way, the necessary condition that a space $V$ is
the ground-state space of some $\K$-local Hamiltonian
is that $V$ is $\cal{K}$-correlated. However,  this 
$\cal{K}$-correlatedness of a space $V$ does not guarantee that
$V$ can be he exact ground-state space of some
$\cal{K}$-local Hamiltonian. 
To understand why this necessary condition may not be sufficient and when the problem
could possibly happen, we link the the spaces satisfying this necessary
condition to faces of the convex body of the local reduced states.
Based on this understanding of convex geometry, 
we then further discuss some methods for constructing the corresponding
$\K$-local Hamiltonians, mainly from physical points of view,
including constructions related to perturbation methods, local
frustration-free Hamiltonians, as well as thermodynamical ensembles.

The perturbation method, combined with the method based on the
frustration-free systems, allows us to identify the $\K$-local
Hamiltonians for all three-qubit states for $\K=\{\{1,2\},\{2,3\}\}$,
except those states which are LU equivalent to $GHZ$-type states. 
In other words, all the extreme points on the corresponding convex
body are exposed in this case. 
Our result then shows that only two out of the three $2$-RDMs are enough
to uniquely determine a three-qubit pure state unless the state is LU
equivalent to a $GHZ$-type state, which is an improvement of the
result given in~\cite{LPW02}.  We also find the $XX$-type Hamiltonians
for $W$ states which are identified in~\cite{XXZ} from other methods.

The method based on an idea of thermal ensembles provides an
alternative and a more physical understanding of the relationship
between $\K$-correlated spaces $\K$-local Hamiltonians, as well as an
numerical method of finding such a $\K$-local Hamiltonian. This
numerical method is based on the continuity principle discussed
in~\cite{Zho08,Zho09a,Zho09b}. And combined with a method of
subsystems which is related to local frustration-free Hamiltonians,
the computational cost may be reduced for some special physical
systems.

One would think the direct way of dealing with the problem of finding
the $\K$-local Hamiltonian for a given $\K$-correlated space is
through a general algorithmic viewpoint. Indeed, this problem can be
straightforwardly formulated in terms of a semi-definite
programming~\cite{BV04}, which can be used to numerically solve this
problem.

However in general, finding a $\K$-local Hamiltonian with a given
$\K$-correlated space $V$ as its exact ground-state space is a very
hard problem.  Theoretically, none of these methods could work if some
$\K$-correlated spaces have a similar property as the point $A$ in
Fig. 1. So it is highly desired to find a theoretical characterization
of those $\K$-correlated spaces which cannot be the ground-state space
of any $\K$-local Hamiltonian, or find a proof to show that such kind
of $\K$-correlated spaces do not really exist.

Also, even if such a $\K$-local Hamiltonian does exist for a
$\K$-correlated space, it is expected that all the methods and
algorithms we have discussed here are not efficient for the general
case. Indeed, one can only expect that each method works well in
certain special cases, as those examples discussed. Future work will
be toward to identify better methods and algorithms for special
situations, especially for $\K$-correlated spaces which are of
interests to quantum information processing, for instance those
resource states for one-way quantum computing~\cite{GE07}. On the
other hand, one would also like to develop methods to identify whether
a space is $\K$-correlated even without finding the corresponding
$\K$-local Hamiltonian.

We hope our work sheds light on the study of relationship between
correlations of quantum states and ground-state spaces of local
Hamiltonians, thus further link the research in both quantum
information science and many-body physics.

\begin{acknowledgments}
  We thank Mary Beth Ruskai, Runyao Duan, and Salman Beige for helpful
  discussions.  JC is supported by NSERC. ZJ acknowledges support from
  ARO and NSF of China (Grant Nos. 60736011 and 60721061). BZ is
  supported by NSERC and CIFAR.  DZ is supported by NSF of China
  (Grant Nos. 10975181 and 11175247) and NKBRSF of China (Grant No. 2012CB922104).
\end{acknowledgments}

\section*{Appendix A. Proof of Observation~\ref{ob:rel}}

To prove the equivalences in Observation~\ref{ob:rel}, we need the
following two lemma.

\begin{lemma}\label{lem:entropy}
  If $\range(\rho_1) \nsubseteq \range(\rho_0)$, there exists $x^\star
  \in (0,1)$ such that
  \begin{equation*}
    S((1-x^\star)\rho_0+x^\star\rho_1) > S(\rho_0).
  \end{equation*}
\end{lemma}

\begin{proof}
  For simplicity, let $\rho_x = (1-x)\rho_0+x\rho_1$. A direct
  calculation gives
  \begin{equation*}
    \begin{split}
      S(\rho_x)-S(\rho_0) & = x \,\bigl(S(\rho_1)-S(\rho_0)+S(\rho_1\|\rho_x)\bigr)\\
      &\;+ (1-x) \, S(\rho_0\|\rho_x).
    \end{split}
  \end{equation*}
  The assumption, $\range(\rho_1) \nsubseteq \range(\rho_0)$, implies
  that $S(\rho_1\|\rho_x) $ can be made arbitrarily large by choosing
  $x$ close to $0$. Therefore, we can find a $x^\star\in(0,1)$
  satisfying $S(\rho_1\|\rho_{x^\star}) > S(\rho_0)-S(\rho_1)$. As
  both terms of the above equation are positive for $x^\star$, we have
  $S(\rho_{x^\star})>S(\rho_0)$.
\end{proof}

\begin{lemma}\label{lem:supp}
  For any quantum state $\rho$, $\range(\rho)\subseteq \range(\trho_k)$.
\end{lemma}

\begin{proof}
  If $\range(\rho)\nsubseteq \range(\trho_k)$, Lemma~\ref{lem:entropy}
  guarantees that there is some number $p\in(0,1)$, such that
  $(1-p)\trho_k + p\rho$ will have larger entropy than $\trho_k$ has.
  This is a contradiction with the definition of $\trho_k$.
\end{proof}

We are now ready to show Observation~\ref{ob:rel}.  We will need to
show for $V=\range(\rho_V)$, where $\rho_V$ is $\K$-correlated, then
for any $\sigma$ supported on $V$, any $\sigma'\in A_{\K}(\sigma)$ is
also supported on $V$. As $\range(\sigma)\subseteq \range(\rho)$, we
can write $\rho = (1-\epsilon) \sigma'' + \epsilon \sigma$ for some
small number $\epsilon$, and $\range(\sigma'')\subseteq V$. Introduce
a new state
\begin{equation}
\label{eq:hatrho}
  \hat{\rho} = (1-\epsilon) \sigma'' + \epsilon \sigma'.
\end{equation}
It is obvious that $\hat{\rho} \in A_{\K}(\rho)$, therefore
$\range(\hat{\rho}) \subseteq \range(\rho)$ by Lemma~\ref{lem:supp} and
$\range(\sigma') \subseteq \range(\hat{\rho}) \subseteq \range(\rho)=V$
where the first inclusion follows from Eq.~(\ref{eq:hatrho}).

\bibliography{StaHam.bib}

\begin{thebibliography}{21}%
\makeatletter
\providecommand \@ifxundefined [1]{%
 \@ifx{#1\undefined}
}%
\providecommand \@ifnum [1]{%
 \ifnum #1\expandafter \@firstoftwo
 \else \expandafter \@secondoftwo
 \fi
}%
\providecommand \@ifx [1]{%
 \ifx #1\expandafter \@firstoftwo
 \else \expandafter \@secondoftwo
 \fi
}%
\providecommand \natexlab [1]{#1}%
\providecommand \enquote  [1]{``#1''}%
\providecommand \bibnamefont  [1]{#1}%
\providecommand \bibfnamefont [1]{#1}%
\providecommand \citenamefont [1]{#1}%
\providecommand \href@noop [0]{\@secondoftwo}%
\providecommand \href [0]{\begingroup \@sanitize@url \@href}%
\providecommand \@href[1]{\@@startlink{#1}\@@href}%
\providecommand \@@href[1]{\endgroup#1\@@endlink}%
\providecommand \@sanitize@url [0]{\catcode `\\12\catcode `\$12\catcode
  `\&12\catcode `\#12\catcode `\^12\catcode `\_12\catcode `\%12\relax}%
\providecommand \@@startlink[1]{}%
\providecommand \@@endlink[0]{}%
\providecommand \url  [0]{\begingroup\@sanitize@url \@url }%
\providecommand \@url [1]{\endgroup\@href {#1}{\urlprefix }}%
\providecommand \urlprefix  [0]{URL }%
\providecommand \Eprint [0]{\href }%
\providecommand \doibase [0]{http://dx.doi.org/}%
\providecommand \selectlanguage [0]{\@gobble}%
\providecommand \bibinfo  [0]{\@secondoftwo}%
\providecommand \bibfield  [0]{\@secondoftwo}%
\providecommand \translation [1]{[#1]}%
\providecommand \BibitemOpen [0]{}%
\providecommand \bibitemStop [0]{}%
\providecommand \bibitemNoStop [0]{.\EOS\space}%
\providecommand \EOS [0]{\spacefactor3000\relax}%
\providecommand \BibitemShut  [1]{\csname bibitem#1\endcsname}%
\let\auto@bib@innerbib\@empty
%</preamble>
\bibitem [{\citenamefont {Nielsen}\ and\ \citenamefont {Chuang}(2000)}]{NC00}%
  \BibitemOpen
  \bibfield  {author} {\bibinfo {author} {\bibfnamefont {M.}~\bibnamefont
  {Nielsen}}\ and\ \bibinfo {author} {\bibfnamefont {I.}~\bibnamefont
  {Chuang}},\ }\href@noop {} {\emph {\bibinfo {title} {Quantum Computation and
  Quantum Information}}}\ (\bibinfo  {publisher} {Cambridge University Press},\
  \bibinfo {address} {Cambridge, England},\ \bibinfo {year} {2000})\BibitemShut
  {NoStop}%
\bibitem [{\citenamefont {Jaynes}(1957)}]{Jay57}%
  \BibitemOpen
  \bibfield  {author} {\bibinfo {author} {\bibfnamefont {E.~T.}\ \bibnamefont
  {Jaynes}},\ }\href {\doibase 10.1103/PhysRev.106.620} {\bibfield  {journal}
  {\bibinfo  {journal} {Phys. Rev.}\ }\textbf {\bibinfo {volume} {106}},\
  \bibinfo {pages} {620} (\bibinfo {year} {1957})}\BibitemShut {NoStop}%
\bibitem [{\citenamefont {Linden}\ and\ \citenamefont {Wootters}(2002)}]{LW02}%
  \BibitemOpen
  \bibfield  {author} {\bibinfo {author} {\bibfnamefont {N.}~\bibnamefont
  {Linden}}\ and\ \bibinfo {author} {\bibfnamefont {W.~K.}\ \bibnamefont
  {Wootters}},\ }\href {\doibase 10.1103/PhysRevLett.89.277906} {\bibfield
  {journal} {\bibinfo  {journal} {Phys. Rev. Lett.}\ }\textbf {\bibinfo
  {volume} {89}},\ \bibinfo {pages} {277906} (\bibinfo {year}
  {2002})}\BibitemShut {NoStop}%
\bibitem [{\citenamefont {Zhou}(2008)}]{Zho08}%
  \BibitemOpen
  \bibfield  {author} {\bibinfo {author} {\bibfnamefont {D.~L.}\ \bibnamefont
  {Zhou}},\ }\href {\doibase 10.1103/PhysRevLett.101.180505} {\bibfield
  {journal} {\bibinfo  {journal} {Phys. Rev. Lett.}\ }\textbf {\bibinfo
  {volume} {101}},\ \bibinfo {pages} {180505} (\bibinfo {year}
  {2008})}\BibitemShut {NoStop}%
\bibitem [{\citenamefont {Linden}\ \emph {et~al.}(2002)\citenamefont {Linden},
  \citenamefont {Popescu},\ and\ \citenamefont {Wootters}}]{LPW02}%
  \BibitemOpen
  \bibfield  {author} {\bibinfo {author} {\bibfnamefont {N.}~\bibnamefont
  {Linden}}, \bibinfo {author} {\bibfnamefont {S.}~\bibnamefont {Popescu}}, \
  and\ \bibinfo {author} {\bibfnamefont {W.~K.}\ \bibnamefont {Wootters}},\
  }\href {\doibase 10.1103/PhysRevLett.89.207901} {\bibfield  {journal}
  {\bibinfo  {journal} {Phys. Rev. Lett.}\ }\textbf {\bibinfo {volume} {89}},\
  \bibinfo {pages} {207901} (\bibinfo {year} {2002})}\BibitemShut {NoStop}%
\bibitem [{\citenamefont {Jones}\ and\ \citenamefont {Linden}(2005)}]{JL05}%
  \BibitemOpen
  \bibfield  {author} {\bibinfo {author} {\bibfnamefont {N.~S.}\ \bibnamefont
  {Jones}}\ and\ \bibinfo {author} {\bibfnamefont {N.}~\bibnamefont {Linden}},\
  }\href {\doibase 10.1103/PhysRevA.71.012324} {\bibfield  {journal} {\bibinfo
  {journal} {Phys. Rev. A}\ }\textbf {\bibinfo {volume} {71}},\ \bibinfo
  {pages} {012324} (\bibinfo {year} {2005})}\BibitemShut {NoStop}%
\bibitem [{\citenamefont {Parashar}\ and\ \citenamefont {Rana}(2009)}]{PR08}%
  \BibitemOpen
  \bibfield  {author} {\bibinfo {author} {\bibfnamefont {P.}~\bibnamefont
  {Parashar}}\ and\ \bibinfo {author} {\bibfnamefont {S.}~\bibnamefont
  {Rana}},\ }\href {\doibase 10.1103/PhysRevA.80.012319} {\bibfield  {journal}
  {\bibinfo  {journal} {Phys. Rev. A}\ }\textbf {\bibinfo {volume} {80}},\
  \bibinfo {pages} {012319} (\bibinfo {year} {2009})}\BibitemShut {NoStop}%
\bibitem [{\citenamefont {Walck}\ and\ \citenamefont {Lyons}(2008)}]{WL08}%
  \BibitemOpen
  \bibfield  {author} {\bibinfo {author} {\bibfnamefont {S.~N.}\ \bibnamefont
  {Walck}}\ and\ \bibinfo {author} {\bibfnamefont {D.~W.}\ \bibnamefont
  {Lyons}},\ }\href {\doibase 10.1103/PhysRevLett.100.050501} {\bibfield
  {journal} {\bibinfo  {journal} {Phys. Rev. Lett.}\ }\textbf {\bibinfo
  {volume} {100}},\ \bibinfo {pages} {050501} (\bibinfo {year}
  {2008})}\BibitemShut {NoStop}%
\bibitem [{\citenamefont {F.~Verstraete}\ and\ \citenamefont
  {Cirac}(2006)}]{VC06}%
  \BibitemOpen
  \bibfield  {author} {\bibinfo {author} {\bibfnamefont {F.}~\bibnamefont
  {F.~Verstraete}}\ and\ \bibinfo {author} {\bibfnamefont {J.~I.}\ \bibnamefont
  {Cirac}},\ }\href {\doibase 10.1103/PhysRevB.73.094423} {\bibfield  {journal}
  {\bibinfo  {journal} {\prb}\ }\textbf {\bibinfo {volume} {73}},\ \bibinfo
  {pages} {094423} (\bibinfo {year} {2006})},\ \Eprint
  {http://arxiv.org/abs/arXiv:cond-mat/0505140} {arXiv:cond-mat/0505140}
  \BibitemShut {NoStop}%
\bibitem [{\citenamefont {Fannes}\ \emph {et~al.}(1992)\citenamefont {Fannes},
  \citenamefont {Nachtergaele},\ and\ \citenamefont {Werner}}]{FNW92}%
  \BibitemOpen
  \bibfield  {author} {\bibinfo {author} {\bibfnamefont {M.}~\bibnamefont
  {Fannes}}, \bibinfo {author} {\bibfnamefont {B.}~\bibnamefont
  {Nachtergaele}}, \ and\ \bibinfo {author} {\bibfnamefont {R.~F.}\
  \bibnamefont {Werner}},\ }\href {\doibase 10.1007/BF02099178} {\bibfield
  {journal} {\bibinfo  {journal} {Communications in Mathematical Physics}\
  }\textbf {\bibinfo {volume} {144}},\ \bibinfo {pages} {443} (\bibinfo {year}
  {1992})}\BibitemShut {NoStop}%
\bibitem [{\citenamefont {Rockafellar}(1996)}]{Roc96}%
  \BibitemOpen
  \bibfield  {author} {\bibinfo {author} {\bibfnamefont {R.~T.}\ \bibnamefont
  {Rockafellar}},\ }\href@noop {} {\emph {\bibinfo {title} {Convex Analysis}}}\
  (\bibinfo  {publisher} {Princeton University Press},\ \bibinfo {year}
  {1996})\BibitemShut {NoStop}%
\bibitem [{\citenamefont {Liu}(2006)}]{Liu06}%
  \BibitemOpen
  \bibfield  {author} {\bibinfo {author} {\bibfnamefont {Y.~K.}\ \bibnamefont
  {Liu}},\ }\href@noop {} {\bibfield  {journal} {\bibinfo  {journal} {Proc.
  RANDOM}\ ,\ \bibinfo {pages} {438}} (\bibinfo {year} {2006})}\BibitemShut
  {NoStop}%
\bibitem [{\citenamefont {Acin}\ \emph {et~al.}(2001)\citenamefont {Acin},
  \citenamefont {Andrianov}, \citenamefont {Jane},\ and\ \citenamefont
  {R.Tarrach}}]{Acin}%
  \BibitemOpen
  \bibfield  {author} {\bibinfo {author} {\bibfnamefont {A.}~\bibnamefont
  {Acin}}, \bibinfo {author} {\bibfnamefont {A.}~\bibnamefont {Andrianov}},
  \bibinfo {author} {\bibfnamefont {E.}~\bibnamefont {Jane}}, \ and\ \bibinfo
  {author} {\bibnamefont {R.Tarrach}},\ }\href@noop {} {\bibfield  {journal}
  {\bibinfo  {journal} {J. Phys. A: Math. Gen.}\ }\textbf {\bibinfo {volume}
  {34}},\ \bibinfo {pages} {6725} (\bibinfo {year} {2001})}\BibitemShut
  {NoStop}%
\bibitem [{\citenamefont {Chen}\ \emph
  {et~al.}(2011{\natexlab{a}})\citenamefont {Chen}, \citenamefont {Chen},
  \citenamefont {Duan}, \citenamefont {Ji},\ and\ \citenamefont
  {Zeng}}]{CCD+10}%
  \BibitemOpen
  \bibfield  {author} {\bibinfo {author} {\bibfnamefont {J.}~\bibnamefont
  {Chen}}, \bibinfo {author} {\bibfnamefont {X.}~\bibnamefont {Chen}}, \bibinfo
  {author} {\bibfnamefont {R.}~\bibnamefont {Duan}}, \bibinfo {author}
  {\bibfnamefont {Z.}~\bibnamefont {Ji}}, \ and\ \bibinfo {author}
  {\bibfnamefont {B.}~\bibnamefont {Zeng}},\ }\href {\doibase
  10.1103/PhysRevA.83.050301} {\bibfield  {journal} {\bibinfo  {journal}
  {\pra}\ }\textbf {\bibinfo {volume} {83}},\ \bibinfo {pages} {050301}
  (\bibinfo {year} {2011}{\natexlab{a}})},\ \Eprint
  {http://arxiv.org/abs/1004.3787} {arXiv:1004.3787 [quant-ph]} \BibitemShut
  {NoStop}%
\bibitem [{\citenamefont {Chen}\ \emph
  {et~al.}(2011{\natexlab{b}})\citenamefont {Chen}, \citenamefont {Ji},
  \citenamefont {Wei},\ and\ \citenamefont {Zeng}}]{CJW+11}%
  \BibitemOpen
  \bibfield  {author} {\bibinfo {author} {\bibfnamefont {J.}~\bibnamefont
  {Chen}}, \bibinfo {author} {\bibfnamefont {Z.}~\bibnamefont {Ji}}, \bibinfo
  {author} {\bibfnamefont {Z.}~\bibnamefont {Wei}}, \ and\ \bibinfo {author}
  {\bibfnamefont {B.}~\bibnamefont {Zeng}},\ }\href@noop {} {\bibfield
  {journal} {\bibinfo  {journal} {ArXiv e-prints}\ } (\bibinfo {year}
  {2011}{\natexlab{b}})},\ \Eprint {http://arxiv.org/abs/1106.1373}
  {arXiv:1106.1373 [quant-ph]} \BibitemShut {NoStop}%
\bibitem [{\citenamefont {Koma}\ and\ \citenamefont
  {Nachtergaele}(1998)}]{XXZ}%
  \BibitemOpen
  \bibfield  {author} {\bibinfo {author} {\bibfnamefont {T.}~\bibnamefont
  {Koma}}\ and\ \bibinfo {author} {\bibfnamefont {B.}~\bibnamefont
  {Nachtergaele}},\ }\href@noop {} {\bibfield  {journal} {\bibinfo  {journal}
  {Adv. Theor. Math. Phys.}\ }\textbf {\bibinfo {volume} {2}},\ \bibinfo
  {pages} {533} (\bibinfo {year} {1998})}\BibitemShut {NoStop}%
\bibitem [{\citenamefont {Bru\ss{}}\ \emph {et~al.}(2005)\citenamefont
  {Bru\ss{}}, \citenamefont {Datta}, \citenamefont {Ekert}, \citenamefont
  {Kwek},\ and\ \citenamefont {Macchiavello}}]{WXX}%
  \BibitemOpen
  \bibfield  {author} {\bibinfo {author} {\bibfnamefont {D.}~\bibnamefont
  {Bru\ss{}}}, \bibinfo {author} {\bibfnamefont {N.}~\bibnamefont {Datta}},
  \bibinfo {author} {\bibfnamefont {A.}~\bibnamefont {Ekert}}, \bibinfo
  {author} {\bibfnamefont {L.~C.}\ \bibnamefont {Kwek}}, \ and\ \bibinfo
  {author} {\bibfnamefont {C.}~\bibnamefont {Macchiavello}},\ }\href@noop {}
  {\bibfield  {journal} {\bibinfo  {journal} {Phys. Rev. A}\ }\textbf {\bibinfo
  {volume} {72}},\ \bibinfo {pages} {014301} (\bibinfo {year}
  {2005})}\BibitemShut {NoStop}%
\bibitem [{\citenamefont {Zhou}(2009{\natexlab{a}})}]{Zho09a}%
  \BibitemOpen
  \bibfield  {author} {\bibinfo {author} {\bibfnamefont {D.~L.}\ \bibnamefont
  {Zhou}},\ }\href {\doibase 10.1103/PhysRevA.80.022113} {\bibfield  {journal}
  {\bibinfo  {journal} {Phys. Rev. A}\ }\textbf {\bibinfo {volume} {80}},\
  \bibinfo {pages} {022113} (\bibinfo {year} {2009}{\natexlab{a}})}\BibitemShut
  {NoStop}%
\bibitem [{\citenamefont {Zhou}(2009{\natexlab{b}})}]{Zho09b}%
  \BibitemOpen
  \bibfield  {author} {\bibinfo {author} {\bibfnamefont {D.~L.}\ \bibnamefont
  {Zhou}},\ }\href@noop {} {\enquote {\bibinfo {title} {An efficient numerical
  algorithm on irreducible multiparty correlations},}\ }\bibinfo {howpublished}
  {arXiv:0909.3700} (\bibinfo {year} {2009}{\natexlab{b}})\BibitemShut
  {NoStop}%
\bibitem [{\citenamefont {Boyd}\ and\ \citenamefont
  {Vandenberghe}(2004)}]{BV04}%
  \BibitemOpen
  \bibfield  {author} {\bibinfo {author} {\bibfnamefont {S.}~\bibnamefont
  {Boyd}}\ and\ \bibinfo {author} {\bibfnamefont {L.}~\bibnamefont
  {Vandenberghe}},\ }\href@noop {} {\emph {\bibinfo {title} {Convex
  Optimization}}}\ (\bibinfo  {publisher} {Cambridge University Press},\
  \bibinfo {address} {Cambridge, England},\ \bibinfo {year} {2004})\BibitemShut
  {NoStop}%
\bibitem [{\citenamefont {Gross}\ and\ \citenamefont {Eisert}(2007)}]{GE07}%
  \BibitemOpen
  \bibfield  {author} {\bibinfo {author} {\bibfnamefont {D.}~\bibnamefont
  {Gross}}\ and\ \bibinfo {author} {\bibfnamefont {J.}~\bibnamefont {Eisert}},\
  }\href {\doibase 10.1103/PhysRevLett.98.220503} {\bibfield  {journal}
  {\bibinfo  {journal} {Phys. Rev. Lett.}\ }\textbf {\bibinfo {volume} {98}},\
  \bibinfo {eid} {220503} (\bibinfo {year} {2007})}\BibitemShut {NoStop}%
\end{thebibliography}%

\end{document}